\def\bq{\begin{equation}}
\def\eq{\end{equation}}
\def\bqa{\begin{eqnarray}}
\def\eqa{\end{eqnarray}}
\def\bqb{\begin{eqnarray*}}
\def\eqb{\end{eqnarray*}}
\documentclass[12pt]{article}\textwidth 16cm
\usepackage{epsfig}
\def\pr#1#2#3{ Phys. Rev. ${\bf{#1}}$ (#2) #3}

\def\pl#1#2#3{ Phys. Lett. ${\bf{#1}}$ (#2) #3 }

\def\np#1#2#3{ Nucl. Phys. ${\bf{#1}}$ (#2) #3}
\def\zp#1#2#3{ Z. Phys. ${\bf{#1}}$ (#2) #3}

\global\nulldelimiterspace = 0pt

\def\Bsl{\hbox{/\kern-.6700em$B$}} 
\def\Dsl{\hbox{/\kern-.6700em$D$}} 
\def\Wsl{\hbox{/\kern-.6700em$W$}} 

\def\roughly#1{\mathrel{\raise.3ex
    \hbox{$#1$\kern-.75em\lower1ex\hbox{$\sim$}}}}

\def\O{ {\cal O }}

\def\mh2{m^2_H}

\begin{document}
\pagenumbering{arabic}
\thispagestyle{empty}
\hspace {-0.8cm} PM/96-14\\
\hspace {-0.8cm} May 1996\\
\vspace {0.8cm}\\

\begin{center}
{\Large\bf Searches for Clean Anomalous Gauge Couplings} \\
\vspace {0.1cm}
{\Large\bf effects at present and future $e^+e^-$ colliders} \\

 \vspace{1.8cm}
{\large  A. Blondel$^a$, F.M. Renard$^b$, L. Trentadue$^c$ and C.
Verzegnassi$^d$}
\vspace {1cm}  \\
$^a$LPNHE, Ecole Polytechnique IN2P3 CNRS 91128 Palaiseau, France\\
and CERN, F01631 CERN Cedex\\
\vspace{0.2cm}
$^b$Physique
Math\'{e}matique et Th\'{e}orique,
CNRS-URA 768,\\
Universit\'{e} de Montpellier II,
 F-34095 Montpellier Cedex 5.\\
\vspace{0.2cm}
$^c$ Dipartimento di Fisica, Universit\'a di Parma\\
INFN, Gruppo Collegato di Parma, 43100 Parma, Italy\\
\vspace{0.2cm}
$^d$ Dipartimento di Fisica,
Universit\`{a} di Lecce \\
CP193 Via Arnesano, I-73100 Lecce, \\
and INFN, Sezione di Lecce, Italy.\\

\vspace{1.5cm}

 {\bf Abstract}
\end{center}
\noindent
We consider the virtual effects of a general type of Anomalous (triple)
Gauge Couplings on various experimental observables in the process of
electron-positron annihilation into a final fermion-antifermion state.
We show that the use of a recently proposed "$Z$-peak subtracted"
theoretical description of the process allows to reduce substantially
the number of relevant parameters of the model, so that a calculation
of observability limits can be performed in a rather simple way. As an
illustration of our approach, we discuss the cases of future measurements
at LEP2 and at a new 500 GeV linear collider.

\vspace{1cm}

\setcounter{page}{0}
\def\thefootnote{\arabic{footnote}}
\setcounter{footnote}{0}
\clearpage

\section{Introduction}\par

Among the various sources of deviations from the Standard Model (SM),
the one that considers the possibility of anomalous triple gauge boson
couplings (AGC) has been very extensively examined and
discussed in recent years. Starting from the undeniable consideration
that for the $WWZ$ and the $WW\gamma$ couplings no stringent
experimental test of the SM predictions is yet available, several
models have been proposed \cite{1} that would predict, or accomodate,
possible differences from the SM canonical values, leading to observable
effects both in present and in future measurements.\par
On this very last topics, some theoretical debate has occurred,
concentrated on the very relevant question of whether the already
available information from experiments at low energy and on
$Z$-resonance peak could, or could not, be improved by a certain set of
future experiments, in particular by those performable at LEP2, for
this special type of models \cite{2a},\cite{2b},\cite{2c}. 
As a result of long and
interesting discussions, it has been generally recognized that 
if the deviations from the SM are
fully incorporated into a theoretical mechanism that retains the
original $SU(2)\times U(1)$ gauge invariance even at a large scale
$\Lambda$ where the SM looses its validity, the available bounds on the
parameters of such models are "mild". One might expect, therefore, that
future experiments at more powerful machines with a suitable
experimental accuracy would improve the bounds for \underline{all} the
parameters, and that the overall improvement would be automatically
guaranteed by moving to higher and higher energy accelerators. In 
this picture, one would guess that a separate analysis of the final two
boson and two fermion channels would lead to increased bounds for the
complete set of parameters, since some of them would only contribute 
the first channel, while the remaining ones would be mainly determined
by the second one. In practice, the final bosonic channel will be
investigated both at $e^+e^-$ and at $pp$, $p\bar p$ colliders. For the
second one, whose analysis requires one loop electroweak effects, the
requested precision should select the $e^+e^-$ colliders as the only
source of possible information. The combined investigations at the two
types of colliders should then lead to a better and better
determination of \underline{all} the parameters of the model.\par
The aim of our paper is to show that this is not always
necessarily the case. To be more specific, we will show that the
previous expectation will be certainly justified for a special subset
of model parameters, for which the bounds should
indeed monotonically increase with c.m. energy . On
the contrary, other parameters do not seem to enjoy this
property. These are those parameters that contribute the final two
fermion channel and that can be reabsorbed in the definition of
LEP1-SLC measured quantities. In this case, the relative accuracy of
the future $e^+e^-$ colliders is beaten (in a certain sense that will be
illustrated) by Z-peak measurements. In a sense, there
appears to be a natural and easy criterion to distinguish those
parameters whose knowledge can be improved by future accelerators from
those for which this would not be the case.\par
In practice, to make this general discussion more explicit, we shall
need a concrete example. With this aim, we shall resort in this paper
to a specific representative model, that
describes the low energy effects of a certain unknown new physics,
appearing at scale $\Lambda$, by an effective Lagrangian built by
dimension six operators only \cite{2a},\cite{3}. 
We shall stick from now on to
the notations of ref.\cite{3}, and devote the interested reader to that
paper for a much more exhaustive discussion of the main points that we
have tried to summarize here.\par
A first, and apparently purely technical, problem immediately arises if
one fully accepts the philosophy and the framework of ref.\cite{3}.
This is related to the relatively large number
of parameters that the model introduces. To describe a four-fermion
process like that of electron-positron annihilation into
fermion-antifermion at arbitrary energy, four renormalized parameters
are requested at the one loop level. To these quantities, that are
specific of the model, one must also add at the considered level the
unknown Higgs mass and the still not extremely precisely determined top
mass, that introduce a small but not really negligible extra
theoretical error in all the fits that try to fix the values of the
four anomalous couplings. Since the number of adequately precise
experimental measurements at such future (including LEP2)
electron-positron colliders is unavoidably limited, a
\underline{conventional} program of derivation of bounds requires some
care (as it was shown in an excellent way in a recent
publication\cite{4}). At the same time, it appears rather cumbersome to
individuate possible features of experimental effects that would be
characteristic of this model (like a definit sign in some deviation, or
special correlations of effects in different observables) and that would
allow, in case of a visible signal, to differentiate the model in a clean
way from other sources of virtual signals.\par
In this short paper we first show that these
difficulties can be greatly reduced if the
"conventional" theoretical description of the considered process is
abandoned and a different one, recently proposed and denominated
"$Z$-peak subtracted" representation, is utilized. To be more precise,
we shall briefly review in the next Section 2 the main features of this
representation, showing that, as an immediate consequence of adopting
it, only \underline{two} parameters of the considered model (without
extra top or Higgs mass dependence) remain in the theoretical
expressions. As a benefit of this simplification, a much simpler
two-parameter fit to the data will be now performable. In Section 3 we
shall give the results of our analysis for the specific cases of LEP2
and of a future 500 Gev linear collider (NLC).
We shall calculate in a realistic way, that takes into account
the potentially dangerous effects of QED radiation, the effects of the
model on various observables and the limits in the
plane of the two surviving parameters.\par
From our reduction of the number of involved parameters, a
second benefit can be derived since it will now be possible to draw in
a $3-d$ space of effects on three different and suitable observables a
region that will be completely characteristic of this model. If other
competitor models admit a theoretical representation of their effect on
the same observables where also only two parameters are involved, they
will also be associated in the previous space to a certain region.
In Section 4 we shall show that, at least for
the two very general models
of "technicolour" type and of "extra $Z$" type, the corresponding
regions (that we shall call "reservations") would not overlap. This
would allow to identify the AGC model, if a virtual signal were seen,
in a relatively clean way, at least with respect to the two previously
mentioned general competitor models.\par
Having shown with a specific example that a sensible reduction of the
number of AGC parameters is indeed possible, we shall devote the final
Section 5 to a short discussion of the possible generalization of our
approach to more complicated cases. We shall try to give reasonable
arguments in favour of the possibility of a systematic 
classification that
would represent a clean compromise between low energy and higher energy
constraints for this type of theoretical models.

\section{The method}
The theoretical description of the process $e^+e^-\to f\bar f$ (here
$f$ is a general fermion) that we follow in this paper has been fully
illustrated in two previous publications, treating separately the case
of final lepton \cite{6} and quark \cite{7} production. In this Section we
shall only illustrate with one representative example the main features
and consequences of our approach.\par
In the theoretical description of the process commonly used at the one
loop level, the invariant scattering amplitude is written as the sum of
a "Born" term and additional higher order "corrections". The input
parameters of the Born term are by convention $\alpha$ (the electric
charge, measured at \underline{zero} momentum transfer), $M_Z$ (the $Z$
mass) and $G_{\mu}$, the Fermi constant defined by the muon lifetime
and known to a relative accuracy of about 2.$10^{-5}$, practically the
same as that now available for the measurement of the $Z$ mass. The
very high accuracy of the experimental determination of these
parameters, that enter as theoretical input the SM predictions,
guarantees that the latter are not affected by unwanted ambiguities.
This is particularly relevant for the set of high precision
measurements performed with the aim of testing the SM by looking for
extra virtual effects on top of $Z$ resonance, where the available
experimental precision for several observables has now reached values
of a relative few permille. Clearly, in this situation, the replacement
in the starting theoretical expressions of $G_{\mu}$ by a different
input parameter known, for instance, at the level of a relative few
permille would not be a productive move.\par
A priori, this rather quantitative consideration does not necessarily
apply for the situation of possible searches of virtual effects beyond
the SM at future $e^+e^-$ colliders, i.e. at LEP2 and at a 500 GeV NLC.
Here the results of a series of dedicated analyses 
\cite{lep2phys}, \cite{9}
show that the realistic experimental precision to be expected for
several relevant observables will be of the order of a relative few
percent. From a purely pragmatic point of view, replacing $G_{\mu}$ by
a parameter known at the few permille level would not lead, in this
case, to \underline{negative} consequences. It is not difficult to show
that the aforementioned replacement might also lead to interesting
\underline{positive} consequences, for suitable choices of the new
parameter(s). To make this statement more precise, we shall consider in
this paper the illustrative and particularly simple example of the pure
$Z$ contribution to the cross section of the process of $e^+e^-$
annihilation into a couple of muons, at squared total c.m. energy $=
q^2$. At the relevant one loop level the latter can be written as :

\bqa
\sigma_{\mu}^{(1)(zz)} (q^{2}) &=& (\frac{4}{3} \pi \, q^{2})
\left | \frac{\sqrt{2}}{4 \pi} \, g^{2}_{A \ell , 0} \, \right.
\left ( 1 + (1 - 4\tilde{s}^{2}_{\ell} (q^{2}))^{2} \right)\times 
 \nonumber \\
& \times & \frac{G_{\mu} M^{2}_{z}}{[ q^{2} - M^{2}_{z} + i \, M_{z}
\Gamma_{z} (q^{2}) ] } \,
\left ( 1 + \frac{\delta G_{\mu}}{G_{\mu}} + \right.
\left. \left. + {\cal R}e \, \frac{\tilde{A}_{z}(0)}{M^{2}_{z}} -
\tilde{I}_{z} (q^{2}) \right) \right|^{2}
\eqa

\noindent
where $g^2_{Al,0}={1\over4}$ and the remaining quantities are defined
as follows. Denoting as $\tilde{A}_i(q^2,\theta)$
($i=\gamma, Z, \gamma Z$) a certain gauge-invariant combination of
transverse self-energy, vertices and boxes, that we always built
following the Degrassi-Sirlin prescription \cite{10}, and following the
definition

\bq A_{i}(q^{2},\theta) \equiv A_{i}(0,\theta) +
q^{2} F_{i} (q^{2},\theta) \eq

\noindent
$s^2_l(q^2)$ is the result of the integration over $cos\theta$ in the
differential cross section of the quantity

\bq  \tilde{s}^2(q^2,\theta) \equiv s^2_1(1+
\tilde{\Delta}\kappa^1(q^2,\theta)) \eq

\noindent
with

\bq \tilde{\Delta}\kappa^1(q^2,\theta) ={c_1\over s_1}
\tilde{F}^{(lf)}_{\gamma Z} (q^2, \theta)
+{c^2_1\over c^2_1-s^2_1}({\Delta \alpha\over \alpha}-{\Delta G_{\mu}
\over  G_{\mu}}-{\Delta M^2_Z\over M^2_Z}) \eq
\noindent
where $\Delta \alpha, \Delta G_{\mu}, \Delta M^2_Z$ are the shifts
from the bare quantities $\alpha_0,\;G_{\mu 0},M^2_{0Z}$
to the corresponding physical ones and
$s^2_1\equiv 1-c^2_1$,  $s^{2}_{1} c^{2}_{1}
= \frac{\pi \alpha}{\sqrt{2}
G_{\mu}M^{2}_{Z}}$.

\noindent
$\tilde{I}_Z(q^2)$
is the result of the analogous operation on the quantity

\bq \tilde{I}_Z(q^2,\theta) =
{q^2\over q^2-M^2_Z}[\tilde{F}_Z (q^2, \theta)-
\tilde{F}_{Z} (M^2_Z, \theta)]  \eq

The possibility of replacing $G_{\mu}$ by a different parameter in
eq.(1) is provided by the observation that the rigourous equality
holds, that defines the leptonic $Z$-width $\Gamma_l$ :

\bq \Gamma_l = ({\sqrt2 G_{\mu}
M^3_{Z}\over48\pi})[1+\epsilon_1][1+(1-4s^2_l(M^2_Z))^2]
(1+\delta_{QED}) \eq

\noindent
where $\epsilon_1$ is the Altarelli-Barbieri parameter \cite{11} :

\bq \epsilon_1 \equiv {\delta G_{\mu}\over G_{\mu}}+
Re\{{\tilde{A}_Z(0)\over M^2_Z}\}-\tilde{I}_Z(M^2_Z) \eq
\noindent
and $s^2_{eff}(M^2_Z)\equiv \tilde{s}^2_l(M^2_Z)$ is the effective
weak mixing angle measured al LEP1/SLC by means of the leptonic
couplings.

Thus, by properly "subtracting" in eq.(1) the combinations
$\tilde{I}_Z(m^2_Z)$ and  $\tilde{s}^2_l(M^2_Z)$
calculated at the $Z$ peak one can
rewrite eq.(1) in the perfectly identical way :

\bqa
\sigma_{\mu}^{(1)(zz)} (q^{2}) & = &
\left ( \frac{4}{3} \pi \, q^{2} \right ) \, \left [ \frac{3
\Gamma_{\ell}}{M_{z}} \right ]^{2} \times \nonumber \\
& \times & \frac{1}{[(q^{2}-M^{2}_{z})^{2} + M^{2}_{z} \Gamma_{z}^{2}]}
\,
\left[ 1 - 2 R(q^{2}) -
\frac{16 (1-4 s^{2}_{1}) c_{1}s_{1} V(q^{2})}
{[1+(1-4 \tilde{s}^{2}_{\ell}(M^{2}_{z}))^{2}]} \right]
\eqa
\noindent
where :

\bq R(q^2) \equiv \tilde{I}_Z(q^2)-\tilde{I}_Z(M^2_Z) \eq

\bq V(q^2) \equiv Re[ F_{\gamma Z}(q^2)- F_{\gamma Z}(M^2_Z) \eq

We can summarize the results of this operation as follows. At one loop,
$G_{\mu}$ can be "traded" for $\Gamma_l$ and $s^2_{eff}(M^2_Z)$ in the
expression of $\sigma_{\mu}$. As a consequence of this exchange, the
"corrections" $\tilde{I}_Z(q^2)$, $F_{\gamma Z}(q^2)$,
are replaced by two "$Z$-peak subtracted" functions $R,V$
and no other $q^2$-independent one-loop theoretical parameters
(${\delta G_{\mu}\over G_{\mu}}$, ${\tilde{A}_Z(0)\over M^2_Z}$,
${\Delta \alpha\over\alpha}$, ...etc)
survive, since they have all been reabsorbed in the definition of the
two \underline{measured} quantities $\Gamma_l$, $s^2_{eff}(M^2_Z)$.\par
The previous discussion applies to the "pure $Z$" contribution to the
muon cross section. For what concerns the two other contributions of
"pure $\gamma$" and of "$\gamma-Z$" type one easily sees that only one
more "canonical" generalized function $\tilde{\Delta}_{\alpha}(q^2)$
, already subtracted at the
"$\gamma$  peak" and entering the photon term, is required at one loop.
This function is conventionally defined as the result of the
$cos\theta$-integration on the generalized quantity
$\tilde{F}_{\gamma}(0,\theta)-\tilde{F}_{\gamma}(q^2,\theta)$ , as one can
easily understand from the previous discussion, and we shall treat it in the
usual way without extra theoretical tricks. The three functions
$R(q^2)$, $V(q^2)$ and $\tilde{\Delta}_{\alpha}(q^2)$
together with $\Gamma_l$ and
$s^2_{eff}(M^2_Z)$ are thus providing at one loop a full theoretical
description for the electroweak component of the muon cross section.
This conclusion is valid also for the most general observables
(polarized and unpolarized asymmetries) that can be measured in the
final charged lepton channel at future $e^+e^-$ colliders.\par
We are now already in a position to show the practical effects of the
used representation for what concerns the calculations of the effects
of the model of AGC ref.\cite{3} that are considered in this paper.
Although a complete discussion has been already given in Ref.\cite{6},
Section 3, we show here with the purpose of being reasonably
self-contained the example that corresponds again to the $Z$-components
of $\sigma_{\mu}$, eq.(1), and we choose the particularly illustrative
case of the term contained in the second round bracket. By a lengthy
but straightforward calculation of the relevant combinations of
self-energies and vertices that make up the gauge-invariant
combination, one is led to the result :

\bqa
\left [  \left ( \frac{\delta G_{\mu}}{G_{\mu}} +
{\cal R}e \, \frac{\tilde{A}_{z}(0)}{M^{2}_{z}} \right ) -
 \tilde{I}_{z} (q^{2}) \right]^{(AGC)} =
\left[  \left ( \frac{-2M^{2}_{W}}{g^{2} \Lambda^{2}} \, f^{r}_{\phi ,
1}
\right ) - 8 \pi \alpha \, \frac{q^{2}}{\Lambda^{2}} \,
\left ( \frac{c^{2}_{1}}{s^{2}_{1}} \, f^{r}_{DW} +
\frac{s^{2}_{1}}{c^{2}_{1}} \, f^{r}_{DB} \right )  \right]
\eqa

A glance to eq.(11) shows that it contains three of the four
renormalized parameters of the model, defined in ref.\cite{3}
as $f^r_{\phi,1},f^r_{DW},f^r_{DB}$.
In the $Z$-peak subtracted representation, eq.(8), the
term eq.(11) is replaced by the subtracted functions $R(q^2)$, whose
expression in the model is :

\bqa
R^{(AGC)}(q^{2}) = 8 \pi \alpha \, \frac{(q^{2}-M^{2}_{z})}{\Lambda^{2}}
\,
\left[ \frac{c^{2}_{1}}{s^{2}_{1}} \, f^{r}_{DW} +
\frac{s^{2}_{1}}{c^{2}_{1}} \, f^{r}_{DB} \right ]
\label{XL}
\eqa

As one sees,  $R(q^2)$ retains only two of the parameters,
i.e. $f^r_{DW},f^r_{DB}$.
The simple reason for this is that the third parameter $f^r_{\phi,1}$
has been reabsorbed in the measured expression of
$\Gamma_l$. Only the two parameters that contribute the
\underline{non constant} part of $\tilde{F}_Z(q^2)$
survive in the subtraction
procedure.\par
It is rather easy to show that the same feature characterizes the
expressions of the two extra subtracted functions $V(q^2)$ and
$\tilde{\Delta}_{\alpha}(q^2)$

\bqa
\tilde{\Delta}^{(AGC)}_{\alpha} (q^{2}) =
-8 \pi \alpha \, \frac{q^{2}}{\Lambda^{2}} \,
\left [ f^{r}_{DW} + f^{r}_{DB} \right ]
\eqa
\bqa
V^{(AGC)}(q^{2}) = 8 \pi \alpha \, \frac{(q^{2}-M^{2}_{z})}{\Lambda^{2}}
\,
\left[ \frac{c_{1}}{s_{1}} \, f^{r}_{DW} -
\frac{s_{1}}{c_{1}} \, f^{r}_{DB} \right ]
\label{XLI}
\eqa

\noindent
and
that the \underline{same} two parameters will appear, in different
linear combinations, in all the three cases i.e. in all the observables
of the final charged lepton channel. This already remarkable fact can
be actually generalized to any observable of a final hadronic channel
generated by the five light ($u,d,s,c,b$) quarks
\cite{7}. The reason that makes this useful simplification possible is
the fact that in this specific model the contribution to
\underline{vertices} are of universal type for massless quarks. 
The only difference with
respect to the leptonic case will be that now the new "effective" Born
approximation will contain hadronic $Z$ width and asymmetries, measured
on top of $Z$ resonance.\par
The final point that has been investigated inRefs.\cite{6} and 
\cite{7} is that of whether the replacement of $G_{\mu}$ with
the set of $Z$ peak observables does not introduce dangerous
theoretical errors. The answer is that, at the expected experimental
accuracy of LEP2 and NLC \cite{lep2phys}, \cite{9} this replacement is
harmless. In conclusion we are in a position to perform a detailed
analysis of the effect of the considered model on the possible
realistic observables. With this aim, for sake of
completeness we list below approximate expressions of 
the various quantities in the
model (the complete and rigorous expressions can be found in
ref.\cite{6},\cite{7}) valid at LEP2, NLC energies.\par
The muonic cross section:

\bqa  \sigma_{\mu}(q^2)&=&\sigma^{Born}_{\mu}(q^2)\bigm\{1+{2\over
\kappa^2(q^2-M^2_Z)^2+q^4}[\kappa^2(q^2-M^2_Z)^2
\tilde{\Delta}\alpha(q^2)\nonumber\\
&&-q^4(R(q^2)+{1\over2}V(q^2))]\bigm\} \eqa
\noindent
where $\kappa\equiv{\alpha M_Z\over3\Gamma_l}\simeq2.64$ and
\bq
\sigma^{Born}_{\mu}(q^2)= {4\pi\alpha^2\over3q^2}
[{ q^4+\kappa^2(q^2-M^2_Z)^2\over\kappa^2(q^2-M^2_Z)^2}]      \eq

\noindent
The muonic forward-backward asymmetry:

\bqa  A_{FB,\mu}(q^2)&=&A^{Born}_{FB,\mu}(q^2)\bigm\{1+
{q^4-\kappa^2(q^2-M^2_Z)^2
\over\kappa^2(q^2-M^2_Z)^2+q^4}[
\tilde{\Delta}\alpha(q^2)+R(q^2)]\nonumber\\
&&+{q^4\over\kappa^2(q^2-M^2_Z)^2+q^4}V(q^2)]\bigm\} \eqa
\noindent
where
\bq
A^{Born}_{FB,\mu}(q^2)= {3q^2\kappa(q^2-M^2_Z)
\over2[ q^4+\kappa^2(q^2-M^2_Z)^2]}      \eq
\noindent

The hadronic cross section:

\bqa  \sigma_{5}(q^2)&=&\sigma^{Born}_{5}(q^2)\bigm\{1+
[{2(q^2-M^2_Z)^2\over 0.81q^4+(q^2-M^2_Z)^2}][\tilde{\Delta}\alpha(q^2)]
\nonumber\\
&&-[{0.81q^4\over 0.81q^4+(q^2-M^2_Z)^2}][2R(q^2)+1.1V(q^2)]\nonumber\\
&&+[{0.06q^2(q^2-M^2_Z)\over 0.81q^4+(q^2-M^2_Z)^2}]
[\tilde{\Delta}\alpha(q^2)-R(q^2)-12.33V(q^2)]\bigm\}  \eqa

\noindent
where

\bqa 
\sigma^{Born}_{5}(q^2)&\simeq&[N^{(QCD)}_Q{44\over27}
{\pi\alpha^2\over q^2}]+[{12\pi q^2\over
[(q^2-M^2_Z)^2+M^2_Z\Gamma^2_Z]}[{\Gamma_l\over M_Z}]
[{\Gamma_5\over M_Z}]\nonumber\\
&\simeq&{44\pi\alpha^2\over9q^2}[1+0.81{q^4\over(q^2-M^2_Z)^2}]
\eqa
\noindent
The b quark production cross section:

\bqa  \sigma_{lb}(q^2)&=&\sigma^{Born}_{lb}(q^2)\bigm\{1+
[{2(q^2-M^2_Z)^2\over 2q^4+(q^2-M^2_Z)^2}]\tilde{\Delta}\alpha(q^2)
-[{4q^4\over 2q^4+(q^2-M^2_Z)^2}][R(q^2)]\nonumber\\
&&-[{q^2[2q^2+1.4(q^2-M^2_Z)]\over 2q^4+(q^2-M^2_Z)^2}]
[V(q^2)]\bigm\}  \eqa

\noindent
where
\bqa
\sigma^{Born}_{lb}&&\simeq [N^{QCD}_b{4\pi\alpha^2\over27q^2}] +
{12\pi q^2\over((s-M^2_Z)^2+M^2_Z\Gamma^2_Z)}[{\Gamma_l\over M_Z}]
[{\Gamma_b\over M_Z}]\nonumber\\
&&\simeq[{4\pi\alpha^2\over9q^2}]
[{ 2q^4+(q^2-M^2_Z)^2\over(q^2-M^2_Z)^2}]
\eqa
\noindent
(a negligible $\gamma Z$ interference term has not been written).

The b quark forward-backward asymmetry:

\bqa  A_{FB,b}(q^2)&=&A^{Born}_{FB,b}(q^2)\bigm\{1+
[{2.27q^2(q^2-M^2_Z)\over 2.27q^2(q^2-M^2_Z)+0.27q^4}]
-{2(q^2-M^2_Z)^2\over 2q^4+(q^2-M^2_Z)^2}]\tilde{\Delta}\alpha(q^2)
\nonumber\\
&&-[{2.27q^2(q^2-M^2_Z)+0.54q^4\over 2.27q^2(q^2-M^2_Z)+0.27q^4}
-{4q^4\over 2q^4+(q^2-M^2_Z)^2}][R(q^2)]\nonumber\\
&&+[{1.4q^2(q^2-M^2_Z)\over  2q^4+(q^2-M^2_Z)^2}-
{3.1q^4\over 2.27q^2(q^2-M^2_Z)+0.27q^4}][V(q^2)]\bigm\} 
\eqa
\noindent
where
\bqa
A^{Born}_{FB,b}\simeq \sigma^{Born}_{FB,b}/\sigma^{Born}_{lb}
\eqa
\noindent
Using
\bqa
\sigma^{Born}_{FB,b}(q^2)&&\simeq {12\pi q^2\over(s-M^2_Z)^2
+M^2_Z\Gamma^2_Z}[{\Gamma_l\over M_Z}][{\Gamma_b\over M_Z}]
{4\tilde{v}_l\tilde{v}_b\over(1+\tilde{v}^2_l)(1+\tilde{v}^2_b)}
\nonumber\\
&&+({8\pi\over3}){q^2-M^2_Z\over(s-M^2_Z)^2+M^2_Z\Gamma^2_Z}
\alpha(0)\sqrt{{\Gamma_l\over M_Z(1+\tilde{v}^2_l)}}
\sqrt{{N^{QCD}_b\Gamma_b\over M_Z(1+\tilde{v}^2_b)}}
\eqa
\noindent
with $\tilde{v}_l$, $\tilde{v}_b$ given by
$\tilde{v}_f=1-4|Q_f|s^2_f(M^2_Z)$, $s^2_f(M^2_Z)$ being effective
quantities measured in LEP1/SLC experiments at Z peak through
suitable asymmetries as explained in ref.\cite{7},
and eq.(22) for $\sigma^{Born}_{lb}$, one obtains :

\bq  A^{Born}_{FB,b}(q^2)={3\over4}[{ 2.27q^2(q^2-M^2_Z)+0.27q^4
\over 2q^4+(q^2-M^2_Z)^2}]  \eq

In eq.(15-26), as one can guess, the first bracket in the r.h.s.
represents what we could call the Z-peak subtracted Born representation
in which $G_{\mu}$ has been systematically replaced by LEP1/SLC
measured quantities.

 A few words of comments on the previous expressions are now in order.
In the leptonic channel, we have considered the muon cross section and
forward-backward asymmetry. In the hadronic case 
we have considered the cross
section for five quarks ($u,d,s,c,b$) production $\sigma_5$ and the
$b\bar b$ cross section and forward-backward asymmetry. All these
quantities will be measured at LEP2 and NLC. 
Other quantities (in particular polarized lepton and quark
asymmetries) that belong to a more distant possible
experimental phase have not been considered. The final $t\bar t$
channel has also not been investigated. In this case, in which the
quark mass plays an important role, an analysis of anomalous gauge
couplings requires a dedicated study, that is beyond the purpose of
this paper.
In the various
expressions, that have been written at variable c.m. energy
$\sqrt{q^2}$, we have only retained those terms that are numerically
relevant in the starting SM expressions and added the AGC shifts only
where it could make experimental sense.\par
In order to perform a rigorous calculation of effects we shall now take
into account in a realistic way the role of the potentially dangerous
QED radiation. From the \underline{convoluted} effects of the model we
shall then derive rigorous bounds on the two surviving parameters. This
will be done in a full detail in the next Section 3 for the specific
case of measurements at LEP2 and NLC.

\section{Derivation of the bounds at LEP2}
\subsection{Calculation of the convoluted effects of the considered AGC
model}

Whenever a virtual (and possibly small) effect has to be
measured and identified, an accurate knowledge of the influence
on the various observables of the QED radiation, that always appears
in any process where charged particles are involved, becomes
unavoidable if a realistic analysis has to be performed.
In fact, as it has been observed several times, the emission of either hard
or soft photons can alter dramatically the shape
and the size of the relevant quantities. In those cases where an
enhancement is produced, a corresponding dilution of a small virtual effect
will be generated, that might reduce or even cancel the
possibility of an identification at the given experimental accuracy.
In order to restore a research program that aims to identify these
virtual effects, it becomes compulsory to take into account with
adequate precision the modification introduced by QED radiation.

In practice initial state radiation
is by far the most relevant part of the QED modifications \cite{LEP1}.
As a consequence of such an emission, soft or hard photons will be
radiated and the available energy will be correspondingly reduced.
If the considered energy range is close to the mass of a resonance, 
the possible dangerous
effect would be a return to the resonance peak, resulting into
obvious and dramatic enhancements of the cross-sections.
To avoid this possibility a proper elimination
of the unwanted radiative return has to be implemented.

The method that we shall follow to evaluate the effects 
of the QED radiation
is the one that uses the so called structure function approach.
The details of the method have been discussed at length in a number
of previous references \cite{stru} and we shall not discuss them here.
In our case, we shall only be interested in unpolarized cross-sections
and forward-backward asymmetries. For these quantities the relevant
theoretical formulae for the general case of production of a final
fermionic $f\bar{f}$ pair can be simply written as follows:

\bq 
\sigma_f (q^2) = \int d x_1 \, d x_2
\, D^e(x_1,q^2) D^{\bar e}(x_2,q^2) \sigma_{0f} 
\left( (1 - x_1 x_2) q^2 \right)
\Theta({\rm cuts}),
\eq
\noindent
where $\sigma_{0f}$ is the lowest order kernel cross-section taken
at the energy scale reduced by the emission of photons,
$D^{e(\bar e)}(x,q^2)$ is
the electron(positron) structure function and 
$\Theta({\rm cuts})$ reproduces
the experimental conditions under which the radiative return will be
evaluated. In order to take into account both soft 
and hard photon emission,
we will use for $D(x,q^2)$ the expression given in 
Ref.\cite{on} by solving,
at the two-loop level, the Lipatov-Altarelli-Parisi evolution equation
in the non-singlet approximation.

An analogous, slightly different expression can be written for a general
unpolarized forward-backward asymmetry. For a final 
$f\bar{f}$ state this reads:

\bq  A_{FB}(q^2) \simeq {1\over \sigma_T(q^2)} \int^1_{z_0} dz
{4z\over1+z^2}H(z)
[\sigma^0_{F}(zq^2)-\sigma^0_{B}(zq^2)], \ \ \ z_0\geq {4m^2_f\over
q^2} \eq

\noindent
where the detailed expression of the radiator 
$H(z)$ can be found
e.g. in Ref.\cite{LEP1}.

In order to perform an explicit calculation, we have proceeded in the
following way. We have firstly written down 
approximate expressions of the
various lowest order kernels that appear in eqs.(27)(28).
Our philosophy has been the
one of writing simple analytic formulae that contain 
the bulk of the Standard
Model expression. With this purpose we have tried 
as a first step to use
our "effective" Born approximation that can be red 
from equations (15)-(26),
first brackets on the r.h.s. . In addition to this, we have
systematically retained
the important one-loop contributions coming from the 
redefinition of the
electric charge $\tilde{\Delta_{\alpha}(q^2)}$. For the 
latter we have only
included the self-energy fermionic contribution. 
For this term an analytic
formula has been given at variable $q^2$ by 
using as normalization
the previous calculation performed at $q^2=M_Z^2$ \cite{burk}. 
We have checked that
the resulting expressions for a range of $q^2$ values 
that belongs to the LEP2
energy region, i.e. from $\sqrt{q^2}=140GeV$ to 
$\sqrt{q^2}\simeq 200GeV$ reproduce the rigorous one-loop 
result of the program
$TOPAZ0$ \cite{topaz} to better that one-percent 
which is certainly enough at
the expected LEP2 level of accuracy \cite{lep2phys}.

Having checked the validity of our kernel expression, 
we have then calculated
the convoluted quantities using eqs.(21)(22). Once 
again, as a cross-check,
we have compared our results with those of $TOPAZ0$, 
under the same conditions
on energy and cuts, and found an agreeement to better than one-percent.

The calculation of the convoluted AGC effects has been finally performed
using the expressions of the shifts due to the model on the subtracted
 quantities
$\tilde{\Delta_{\alpha}},R,V$ given in eqs.(12)-(14) and 
implementing them in a
dedicated numerical program \cite{tressi}.

The results of the calculation, that have been performed 
choosing for the
experimental cuts the value $z=1-x_1x_2=0.65$ and fixing 
conventionally the scale
parameter of the model, $\Lambda$, at one GeV, are shown 
in Table 1 at different
values of $f_{DW}\;f_{DB}$ for three variables, i.e. 
$\sigma_{\mu},A^{FB}_{\mu}$
and $\sigma_5$ ($\bar \delta$ representing the relative shifts). 
We have also calculated the effect on the remaining
unpolarized
variables $\sigma_b$ and $A^{b}_{FB}$. However, as we shall discuss
in the second half of the Section this model is not 
able to produce observable
effects on these quantities at LEP2 under realistic 
experimental conditions,
and for this reason we have not shown the corresponding 
numbers in the table.

\begin{table}[htb]\centering
\caption[AGC]
{\sl AGC effects on observables.}
\vspace*{2mm}
\begin{tabular} {|c|c|c|c|c|}
\hline
 \hline
&&&&\\
{$f_{DW}$} &
{$f_{DB}$} &
{$\bar\delta\sigma_{\mu}$} &
{$\bar\delta A_{FB,\mu}$} &
{$\bar\delta\sigma_5$} \\
&&&&\\
\hline
\hline
  -1 & -4 & 0.051 & -0.0062 & 0.028    \\[0.1cm]\hline
  -1 & -2 & 0.034 & 0.013& 0.023  \\[0.1cm]\hline
  -1 & 0 & 0.016 &0.032& 0.017\\[0.1cm]\hline
  -1  & 2 & -0.00055  & 0.053&  0.012\\[0.1cm]\hline
  -1 & 4 &-0.018  & 0.074& 0.0070   \\[0.1cm]\hline
 0 & -4 & 0.034 &  -0.038& -0.0044\\[0.1cm]\hline
0 & -2 & 0.017&-0.019& -0.0096  \\[0.1cm]\hline
0 & 0 & 0 & 0 & -0.015\\[0.1cm]\hline 
 0  & 2 & -0.017 & 0.020 & -0.020\\[0.1cm]\hline  
 0  & 4 &-0.034  & 0.041 & -0.025\\[0.1cm]\hline
1  & -4 & 0.018 & -0.071 & -0.037\\[0.1cm]\hline
1 & -2  & 0.00055& -0.053 & -0.042\\[0.1cm]\hline
1 & 0 &-0.016 & -0.033 & -0.047 \\[0.1cm]\hline
  1  & 2 & -0.034  & -0.014 & -0.052 \\[0.1cm]\hline
 1  & 4 &-0.051 & 0.0069 & -0.057 \\[0.1cm]
\hline \hline
\end{tabular}
\label{AGC}
\end{table}

As one sees from Table 1, the convoluted shifts can be, for a sizeable
range of the values of the parameters, of the order 
of a relative few percent.
These would be visible at LEP2 in the next future 
configuration $\sqrt{q^2}=
175 GeV$ with an integrated luminosity of $500 pb^{-1}$, since the
relative experimental accuracy for all these 3 observables would be
about one percent, as shown in details in the numerical tables of
ref.\cite{lep2phys}.
From now on we shall
therefore concentrate our attention on this 
experimental situation. In the
second half of the Section we shall discuss 
the bounds on $f_{DW}$ and $f_{DB}$
that will be correspondingly derived.

\subsection{Derivation of the limits on the AGC parameters}

In the derivation of bounds on the two residual parameters
$f_{DW}\,f_{DB}$
we used the five experimental quantities
of equations (15)-(26). At LEP2, in the chosen configuration,
their relevance will be fixed by the realistically expected
experimental conditions, that will priviledge some observables
with respect to the other ones.
In order to fully understand
this important feature, we discuss at this point some experimental
details.

A preliminary question concerns the choice of the most suitable
event selection. In addition to experiment-dependent cuts on
final state particle angles and momentum, there is a degree
of freedom in the choice of the minimum visible invariant mass
of the fermion anti-fermion pair that is produced, or, equivalently,
in the value of the maximum fraction of center-of-mass energy
$ x_{max}=1-x_1x_2 $  carried away by initial state radiation.

Originally, the various cross-sections were evaluated using the
programm TRESSI at $\sqrt{q^2} = 175$~GeV for a value
of $x_{max}=0.65$.
By varying the cut $x_{max}$, we found that the best sensitivity
for all the investigated cross-sections occurred rather at a value of
$x_{max}\simeq 0.4$, that corresponds to a minimum fermion
invariant mass of 135~GeV.
Fig. 1 shows the typical sensitivity as a function of $x_{max}$,
for the most relevant case of $\sigma_{hadrons}$.
The dependence is though rather flat from
$x_{max}=0.1$ to $0.65$.
From here on, we shall work at the optimal point $x_{max}=0.4$.
Of course, the exact choice will be dictated by specific experimental
considerations.

\begin{figure}[htb]
\vspace*{-.5cm}
\begin{center}
\mbox{
 \epsfig{file=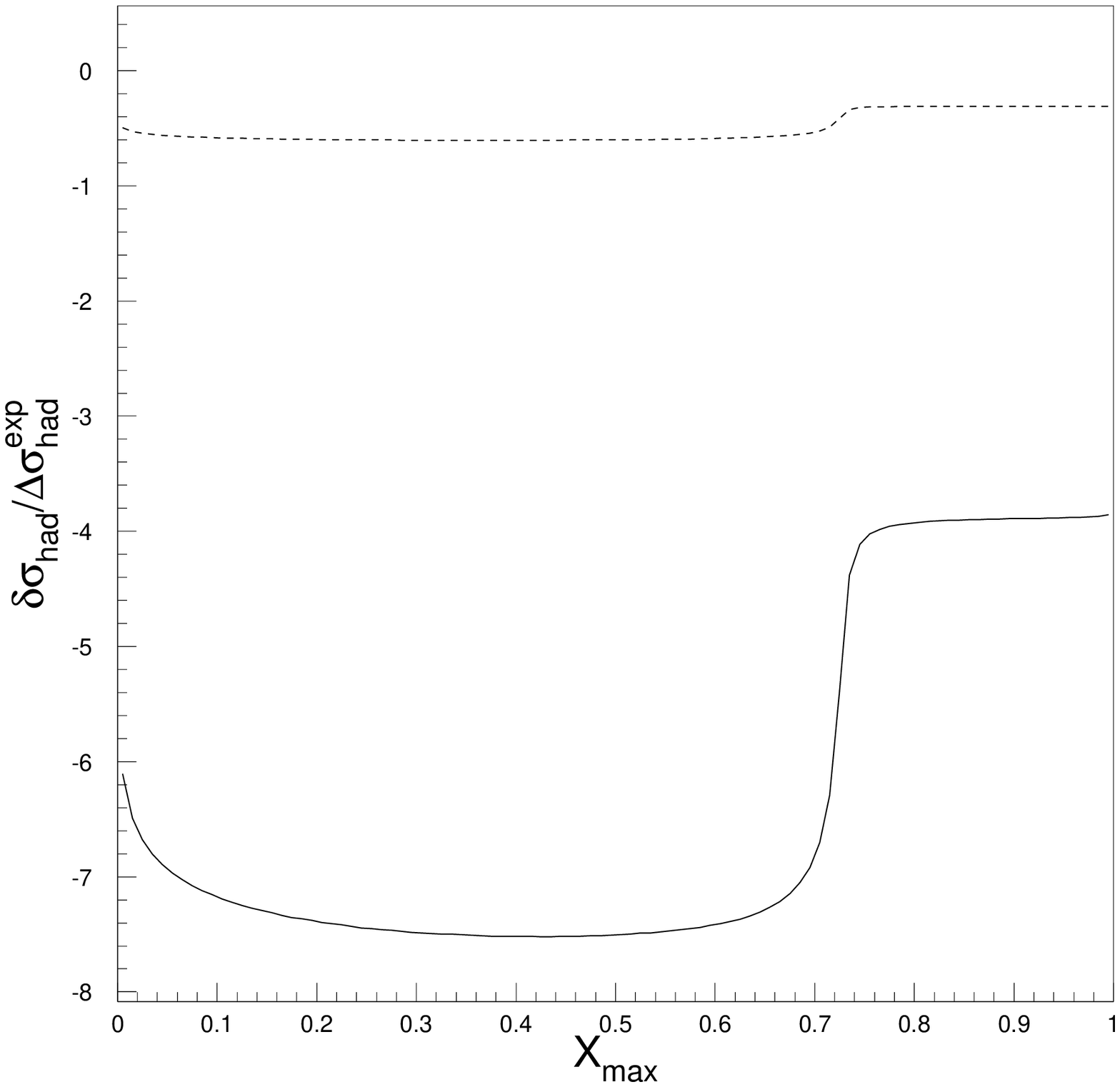,width=150mm}
}
\end{center}
\caption[Sensitivity of Hadronic]
{{\sl
Sensitivity of the hadronic cross-section to $f_{DW}$ (full line)
and $f_{DB}$ (dashed line), as a function of the fraction $x_{max}$ of
center-of-mass energy carried away by initial state photons.
}}
\label{senslep2}
\end{figure}

In the determination of experimental errors for the
various observables, we have made the following
assumptions. The hadronic detection efficiency was assumed to be
95\%; that for $\mu $ and $\tau$ pairs, 90\%.
for $b\bar{b}$ pairs, 50\%. Systematic errors were assumed
to be smaller than the statistical ones, which are in all cases
larger than 0.4\%, and neglected.
The quoted errors were obtained assuming an exposure of 500 pb$^{-1}$
for each of the four LEP experiments.

Working in this realistic LEP2 experimental picture, we found that the
considered model is in practice unable to affect $\sigma_b$ 
and $A_{FB,b}$. This would not necessarily be true at LEP2
for a different theoretical model. 

The results of our estimates are given in Table~2, that shows for
each observable, the expected value and error.
From an inspection of that table on can see that, a priori,
the most promising quantity is  $\sigma_{hadrons}$ followed
by $A_{FB}^{\mu}$ and
$\sigma_{\mu}$.

The constraints on
$f_{DW}$ and $f_{DB}$ were obtained
from each of these observables first,
then from their combination as follows.
The measurement was assumed to give as central value the SM
result. One standard deviation bands and contour were then
drawn on the  $f_{DW}\,f_{DB}$ plane as shown in Fig.2.
One can see that the main contributors to the overall bounds
are  $\sigma_{hadrons}$  and $A_{FB}^{\mu}$.
This latter quantity is in fact the only one that crosses
in a useful way the band provided by $\sigma_{hadrons}$ .
Numerically, our results can be written as follows:
\begin{eqnarray}
\Delta f_{DW} = \pm 0.13 \\
\Delta f_{DB} = \pm 0.73
\end{eqnarray}
with a negative correlation.\par
The equations (29), (30) and Fig.2 represent one of
the main results of this paper, showing the bounds of the two
surviving AGC parameters that would be derivable at LEP2.

\begin{figure}[htb]
\begin{center}
\mbox{
 \epsfig{file=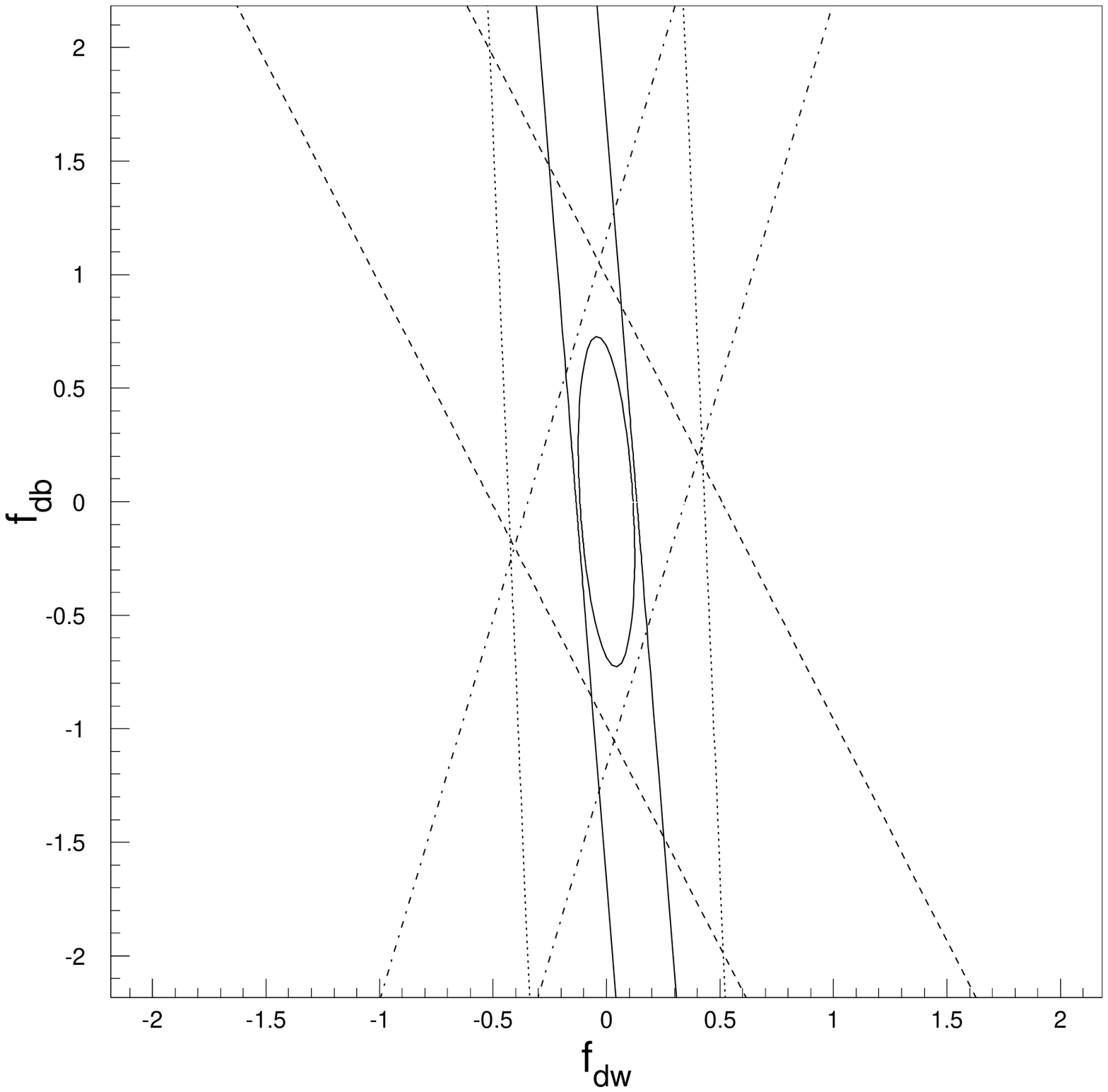,width=150mm}
}
\end{center}
\caption[LEP 2 Bands]
{{\sl
Constraints in the  $f_{DW},~f_{DB}$ plane resulting from
the measurements at LEP2 (4 experiments, 500 pb$^{-1}$ each)
of  $\sigma_{hadrons}$  (full lines)
$\sigma_{b\bar{b}}$ (dotted lines),
$\sigma_{\mu}$ (dashed lines),
and  $A_{FB}^{\mu}$ (dash-dotted lines). The ellipse represents
the one standard deviation (39\% C.L.) constraint resulting
from the combination of the four above measurements.
}}
\label{boundslep2}
\end{figure}

We have also examined the precision of a similar
analysis for a possible New
$e^+e^-$ Linear Collider  (NLC)
at 500~GeV center-of-mass energy with an
integrated luminosity of 20~fb$^{-1}$.
Using the same programme TRESSI to evaluate cross-sections
and asymmetries, and using the available information on
experimental conditions, we have found from the analysis
of
$\sigma_{hadrons}$, $\sigma_{\mu}$ and  $A_{FB}^{\mu}$,
the bounds illustrated in Fig.3.
\begin{figure}[htb]
\begin{center}
\mbox{
 \epsfig{file=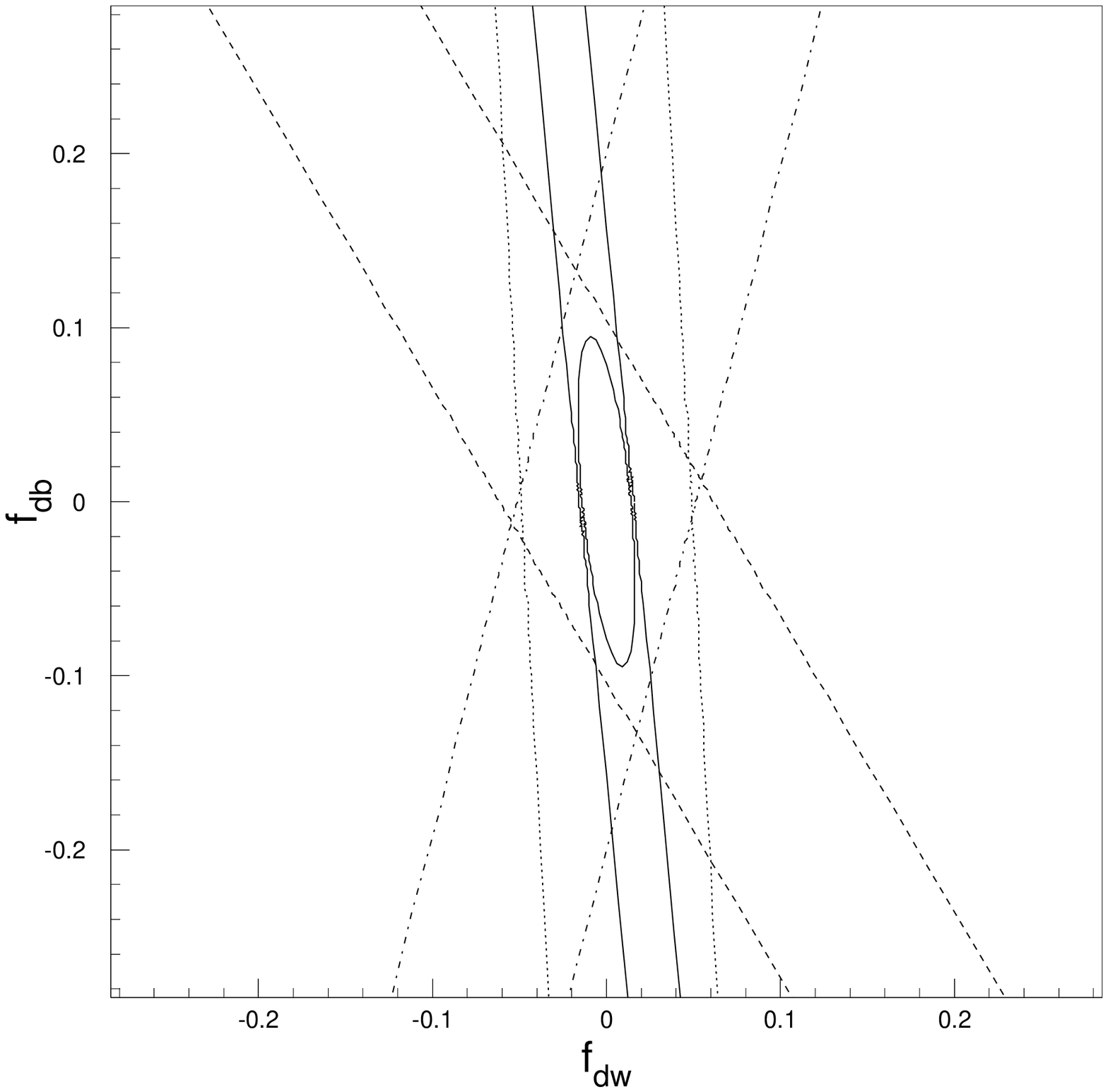,width=150mm}
}
\end{center}
\caption[LEP 2 Bands]
{{\sl
Constraints in the  $f_{DW},~f_{DB}$ plane resulting from
the measurements at NLC  (1 experiments, 20 fb$^{-1}$)
of  $\sigma_{hadrons}$  (full lines)
$\sigma_{b\bar{b}}$ (dotted lines),
$\sigma_{\mu}$ (dashed lines),
and  $A_{FB}^{\mu}$ (dash-dotted lines). The ellipse represents
the one standard deviation (39\% C.L.) constraint resulting
from the combination of the four above measurements.
}}
\label{boundslep500}
\end{figure}

The errors on
$f_{DW}\,f_{DB}$ become:
\begin{eqnarray}
\Delta f_{DW} = \pm 0.016 \\
\Delta f_{DB} = \pm 0.095
\end{eqnarray}
which is one order of magnitude more precise than at LEP2,
a fact that calls for a comment.
We took a mildly optimistic point of view that the 
experimental errors on the absolute cross-section measurement
would be no larger than at LEP2, e.g.0.25\%.
The dramatic improvement in the bounds is therefore due, in this
case, to our expectation of accurate luminosity measurements at NLC. 
This represents, in our opinion,
a very strong motivation in favour of such a performance.

\begin{table}[htb]\centering
\caption[LEP2 results]
{\sl Observables at LEP2: value, experimental errors, sensitivity
to AGC couplings. Cuts, efficiencies and experimental
 precisions as described in the text.}
\vspace*{2mm}
\begin{tabular} {||c|c c c c ||}
\hline
 \hline
&&&&\\
 $Observable {\cal O}$& Value & exp. error &
${{\partial {\cal O}\over\partial{f_{DW}}  }}$ &
${{\partial {\cal O}\over\partial{f_{DB}}  }}$ \\
&&&&\\
\hline
\hline
&&&&\\
$\sigma_{hadrons}$ (pb) & 28.7    &   0.12 & -0.92   & -0.07 \\
&&&&\\
$\sigma_{b\bar{b}}$ (pb)&  4.7    &   0.07 & -0.16   & -0.007\\
&&&&\\
$\sigma_{\mu}$      (pb)& 4.05    &   0.05 & -0.066  & -0.034\\
&&&&\\
  $A_{FB}^{\mu}$        & 0.58    &   0.01 & -0.019  & +0.006\\
&&&&\\
\hline \hline
\end{tabular}
\label{lep2_sensitivity}
\end{table}

\section{Comparison of the effects of different models}

As an undeniable benefit of our approach, we have been able to perform
in the previous sections
two parameter fits to derive bounds for the two surviving quantities
 $f_{DW}$ and $f_{DB}$. To our knowledge, this is the only available
determination of such a simplicity, that avoids the more elaborate
procedures involved when four parameters (plus the Higgs and top
masses) are simultaneously fitted.\par
To add to this paper a somewhat speculative analysis, we shall
consider the case in which a certain signal of \underline{virtual} type
has been "cleanly" seen e.g. at LEP2 (a completely similar discussion
would apply for NLC). For simplicity, we shall treat this effect in
Born approximation, and shall assume that a reasonably accurate
measurement of the final $\tau$ longitudinal polarization $A_{\tau}$
has been performed (in our previous realistic treatment, we did not
include this measurement since at $\sqrt{q^2}=175 GeV$ it would only
react to rather large values of the parameters).
This possibility would become much more realistic at NLC if
longitudinal polarization were available. In fact, the theoretical
expressions of $A_{\tau}$ and of the longitudinal polarization
asymmetry for final \underline{lepton} production $A^l_{LR}$ are
identical. However, the experimental precision of $A^l_{LR}$ at NLC
would be much higher than that of $A_{\tau}$.\par
For sake of completeness, we write here the theoretical expression of
$A_{\tau}$ that is analogue to our previous eqs.(15)-(26).

\bqa A^l_{LR}(q^2)&=&A^{l,Born}_{LR}(q^2)
\bigm\{1+[{\kappa(q^2-M^2_Z)
\over\kappa(q^2-M^2_Z)+q^2}-{2\kappa^2(q^2-M^2_Z)^2\over
\kappa^2(q^2-M^2_Z)^2+q^4}]
[\tilde{\Delta}\alpha(q^2)\nonumber \\
&&+R(q^2)]
-{4c_1s_1\over v_1}V(q^2) \bigm\} \eqa

\noindent
where
\bq
A^{l,Born}_{LR}(q^2)= {q^2[\kappa(q^2-M^2_Z)+q^2]
\over \kappa^2(q^2-M^2_Z)^2+q^4}A(M^2_Z)      \eq
\noindent
$A(M^2_Z)$ being the LR asymmetry at Z peak directly measured at SLC or
indirectly through $A_{FB,\mu}$ or $A_{\tau}$ at LEP1.\par

 Adding to this
observable the muon cross section and asymmetry, one has three
independent leptonic quantities and two surviving anomalous parameters.
This means that the \underline{shift} on $A_{\tau}$ will be given in
terms of those on $\sigma_{\mu}$, $A_{FB,\mu}$ in a way that will not
depend on $f_{DW}$ and $f_{DB}$. Otherwise stated, it will be possible
to draw a certain region in the $3d$ space of the shifts
$\delta A_{\tau}$, $\delta\sigma_{\mu}$, $\delta A_{FB,\mu}$ 
that will be
characteristic of the model and that we shall call "AGC reservation at
LEP2, NLC".\par
Identical conclusions would be derivable for any model whose effects
on the three previous observables may be expressed by two parameters
only. In previous references \cite{6}, \cite{7} we 
considered two specific such cases,
i.e. that of a model of "technicolour type" \cite{TC} with two
strong vector and axial-vector resonances, and that of a model with one
extra $Z\equiv Z'$ with the most general couplings to charged leptons.
The corresponding "reservations" can be easily drawn. This has been
done in full details in reference \cite{6}. Here we shall
only show in Fig.4,5 the three different reservations that correspond
to these three models (called AGC, TC and Z') at LEP2.

\begin{figure}[htb]
\begin{center}
\mbox{
 \epsfig{file=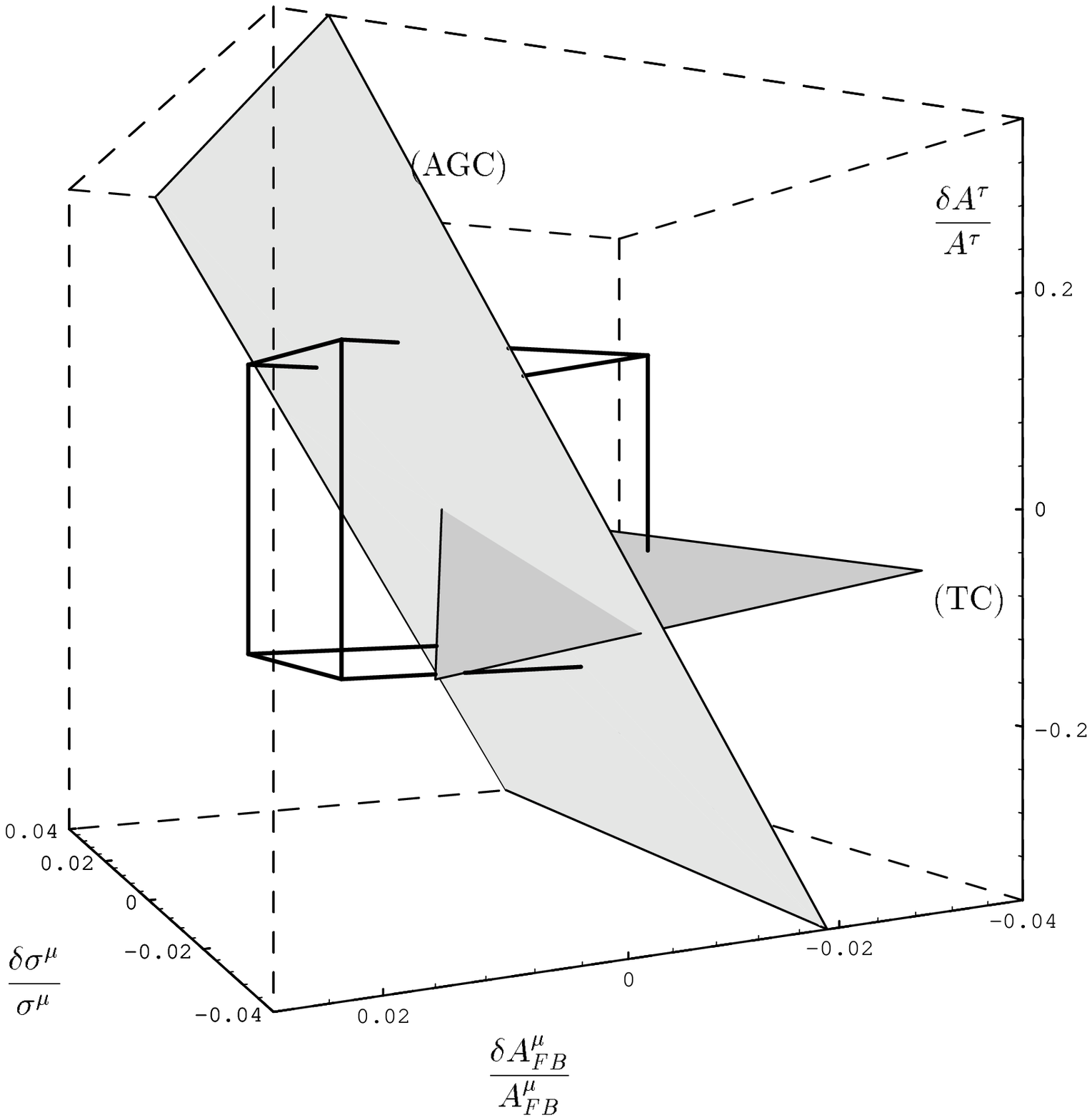,width=150mm}
}
\vspace*{-2cm}
\end{center}
\caption[LEP 2 Bands]
{{\sl Trajectories in the 3-dimensional
space of relative departures from SM for
leptonic and hadronic observables
$\sigma_{\mu}$, $A_{FB,\mu}$, $A_{\tau}$  at a LEP2 energy of 175 GeV
for AGC
models and TC models.\\
 The box represents the unobservable domain
corresponding to a relative accuracy of 1.5 percent for 
$\sigma_{\mu}$, $A_{FB,\mu}$ and 15 percent for $A_{\tau}$.}}
\label{fig4}
\end{figure}

\begin{figure}[htb]
\begin{center}
\mbox{
 \epsfig{file=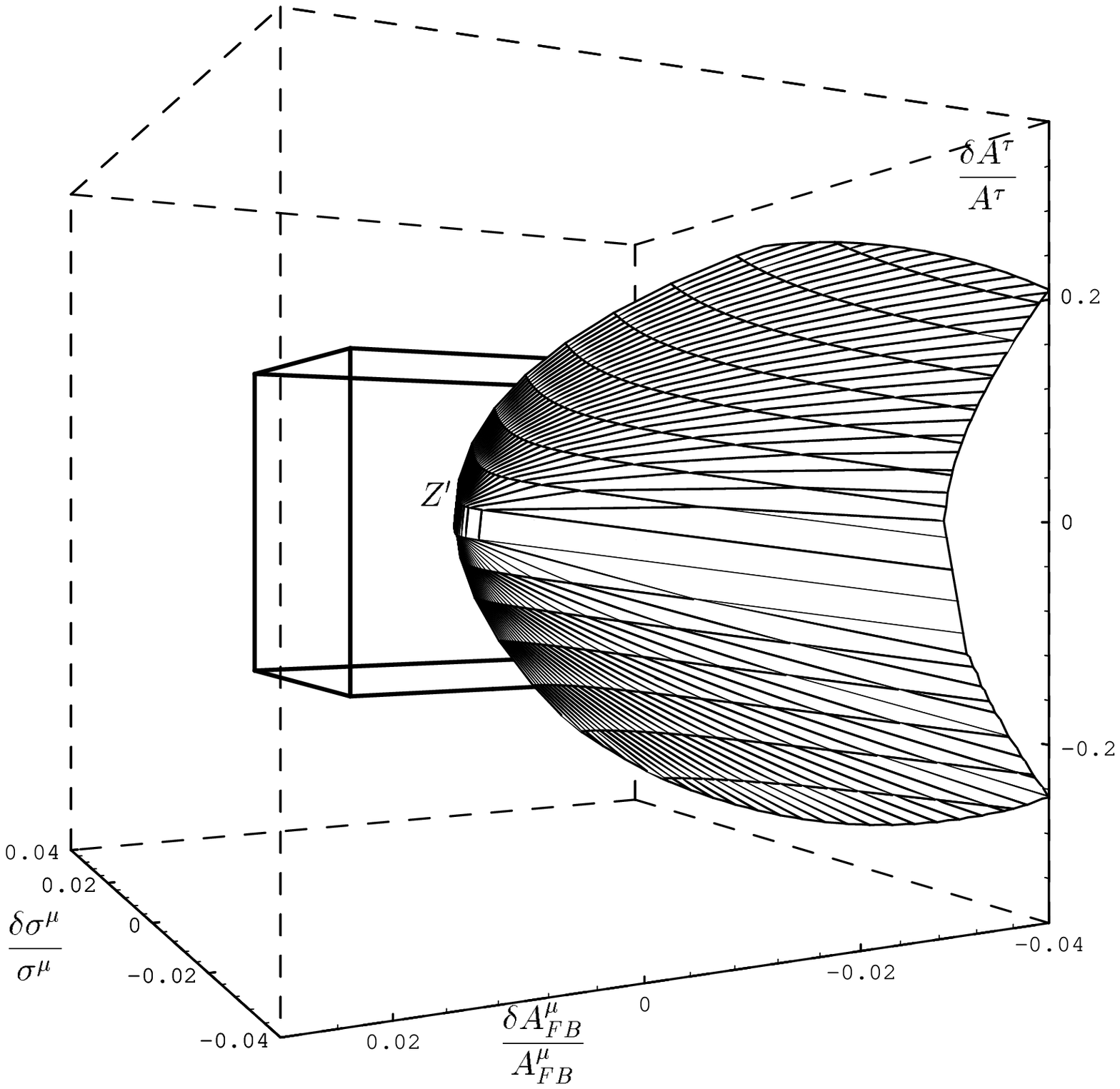,width=150mm}
}
\vspace*{-1.2cm}
\end{center}
\caption[LEP 2 Bands]
{{\sl Trajectories in the 3-dimensional
space of relative departures from SM for
leptonic and hadronic observables
$\sigma_{\mu}$, $A_{FB,\mu}$, $A_{\tau}$  at a LEP2 energy of 175 GeV
for general $Z'$
models.}}
\label{fig5}
\end{figure}

 As one
sees, there is practically no overlapping in the meaningful region of
the shifts space. This allows us to claim that, should a clear virtual
effect manifest itself in the final lepton channel at LEP2, it
would be possible to identify the responsible model within the limited
(but reasonably representative) set of still surviving theoretical
competitors. Our conclusions are obviously made possible by the fact
that the number of involved parameters was reduced to two. Adding
this final discussion to the results obtained in Section 3 we would
therefore state, as claimed in the Introduction, that from our
Z-peak subtracted approach a search of clean effects of a class of
models with anomalous gauge couplings at future $e^+e^-$ colliders
would, indeed, be made possible.\par

\section{Conclusions}

We have shown in this paper that a "Z-peak subtracted" representation
of four fermion (neutral current) processes allows to derive in a
simple way realistic bounds for a reduced number of parameters of
certain general models with Anomalous Gauge Couplings. The parameters
that benefit from this approach are those that contribute the 
\underline{non constant} part of the generalized self-energies 
$\tilde{F}_i (q^2)$, $i=Z, \gamma Z, \gamma$. Other
parameters are reabsorbed in the definition of various quantities
\underline{measured} on the $Z$ peak, that appear as new theoretical
inputs replacing $G_{\mu}$.\par
This conclusion can be reexpressed in a way that represents sort of 
a compromise between previous discussions about the role of
LEP1/SLC measurements with respect to LEP2 investigations \cite{2a},
\cite{2b},\cite{2c},\cite{3}. In
our opinion, it is undeniable that a subset of the "LEP1 blind"
parameters of the model are also "LEP2, NLC final-2-light
fermion channel
blind".
These are precisely those parameters that can be reabsorbed in $Z$-peak
quantities, given their available experimental accuracy and given
\underline{the
realistic expected accuracy} at LEP2, NLC. In the model that we have
considered, these parameters are called $f_{BW}$ and $f_{\Phi,1}$. We
cannot derive for their bounds any improvement when moving
from LEP1/SLC to the LEP2 and NLC \underline{final light fermion
channels}. No direct information should also be expected 
on these parameters from the 
\underline{$WW$ channel}. $\O_{\Phi,1}$ and $O_{DB}$ 
do not generate 3-gauge boson couplings 
($O_{DW}$ and  $O_{BW}$ do generate 3-boson couplings but due to
the available  LEP1 constraints they lie at an 
unobservable level in this channel). 
The $WW$ channel should only be fruitful for
studying the blind
operators $O_{WWW}$, $O_W$ and $O_B$.\par
The previous statements are supposed to be valid for a (neutral
current) four fermion process. Here the Z-peak subtracted
representation can be used. For other types of processes (like for
instance charged current four fermion ones) this prescription cannot be
utilized at least in the present formulation. In such cases, the
conventional representation using $G_{\mu}$ can be used. An example of
this type would be represented by a measurement of the W mass, whose
theoretical expression depends also on the two parameters 
$f_{BW}$, $f_{\Phi,1}$ that cannot be reabsorbed in this case. In 
fact in our opinion, $M_W$ should be used in a separate fit to the AGC
parameters \underline{together} with the various Z-peak data and
considered as another "low energy input".\par 
One might imagine that further information on $f_{BW}$, $f_{\Phi,1}$
would be brought by the
study of 
final $t\bar t$ states. Here, a priori, our subtraction
technique cannot be applied so simply (because the necessary input
$\Gamma _{Zt\bar t}$ does not exist). The fact is, though, that in this
case a (probably) large number of extra $\simeq m^2_t$ parameters would
appear (clearly in a not universal way), and the full analysis would
become much more complicated.\par
To conclude this paper, we have considered
the conventional analysis of ref.\cite{4} where all the four parameters
are retained. This comparison requires some care since the experimental
picture and the computational details utilized there in the fit
are not 
identical with ours. We can still remark that the bounds on
$f_{DW}$, $f_{DB}$ are qualitatively consistent with ours. 
For the remaining two
parameters, we see that, indeed, the relative improvement of
ref.\cite{4} from LEP2 to
NLC is much weaker than that on the remaining two, in agreement with
our expectations. There is an improvement from LEP1 to LEP2 for
$f_{BW}$, $f_{\Phi,1}$ but this should be due, in our opinion, to the
fact that the
information from LEP2 contains also an assumed strongly improved
measurement of $M_W$, which depends effectively, as we said, on
$f_{BW}$, $f_{\Phi,1}$.\par
In principle, our approach could be generalized to models with a larger
number of parameters. For instance, one might consider dimension eight
operators in a model with AGC. Since those
parameters that contribute the non constant component of the
functions $F_i(q^2)$ would survive, in a model like this with
higher dimension operators there would certainly be several ones 
enjoying this property (e.g.
of derivative type). 
Our statement is that our representation would free
the various observables from spurious contributions from parameters
like $f_{BW}$, $f_{\Phi,1}$ that could hide the 
determination of those parameters that are really
effective at high energies, in particular those that would have a
quartic increase $\simeq q^4/\Lambda^4$. 
With a sufficient number of experimental
quantities a complete determination of the meaningful parameters might
then be realistically achieved.

\underline{Acknowledgements}: One of us (L.T.) acknowledge the
hospitality received at the CERN TH division where a part of this work
was carried out.



\begin{thebibliography}{99}



\bibitem{1} K.J.F. Gaemers and G.J. Gounaris,
\zp{C1}{1979}{259}; K. Hagiwara, R. Peccei, D. Zeppenfeld and K.
Hikasa, \np{B282}{1987}{253}.

\bibitem{2a} 
 A. De R\'{u}jula, M.B. Gavela, P. Hernandez 
and E. Masso,\np{B384}{1992}{3}.

\bibitem{2b} 
 C. Grosse-Knetter, I. Kuss and D. Schildknecht,
\zp{C60}{1993}{375}.

\bibitem{2c} 
M. Bilenky, J.L. Kneur, F.M. Renard and
D. Schildknecht, \np{B409}{1993}{22} and
{\bf{B419}} (1994) 240.

\bibitem{3} K. Hagiwara, S. Ishihara, R. Szalapski
and D. Zeppenfeld, \pl{B283}{1992}{353} and \pr{D48}{1993}{2182}.

\bibitem{4} K. Hagiwara, S. Matsumoto and R. Szalapski,
\pl{B391}{1995}{411}.


\bibitem{6} F.M. Renard and C. Verzegnassi, \pr {D52}{1995}{1369}.

\bibitem{TC} J. Layssac, F.M. Renard and C. Verzegnassi, 
\pr{D49}{1994}{2143}.


\bibitem{7} F.M. Renard and C. Verzegnassi, \pr {D53}{1996}{1290}.

\bibitem{lep2phys} Physics at LEP2, CERN 96-01, 
G. Altarelli, T. Sjostrand and F. Zwirner eds.

\bibitem{9} $e^+e^-$ Collisions at 500 GeV: The Physics Potential, DESY
93-123C, p.345 (1993), P.M. Zerwas ed.

\bibitem{10} G. Degrassi and A. Sirlin,
\np {B383} {1992} {73} and \pr {D46} {1992} {3104}.

\bibitem{11}  G. Altarelli, R. Barbieri, F. Caravaglios, \pl {B314}
{1993} {357}.


\bibitem{vertex}  J. Layssac, F.M. Renard and C. Verzegnassi,
\pr{D49}{1994}{3650}.



\bibitem{LEP1} Z Physics at LEP1, CERN 89-08, Vol.1, p.203, G.
Altarelli, R. Kleiss and C. Verzegnassi, eds.

\bibitem{stru} For a review see also: 
O.~Nicrosini and L.~Trentadue,
in {\it Radiative Corrections for $e^+ e^-$ Collisions},
J.~H.~K\"uhn, ed. (Springer, Berlin, 1989), p.~25; in {\it QED Structure
Functions}, G.~Bonvicini, ed., AIP Conf. Proc. No.~201
(AIP, New York, 1990), p.~12; O.~Nicrosini, ibid., p.~73. 

\bibitem{on} O. Nicrosini and L. Trentadue, \pl{B196}{1987}{551},
\zp{C39}{1988}{479}.

\bibitem{burk} H. Burkhardt, F. Jegerlehner, G. Penso and C.
Verzegnassi, \zp{C43}{1989}{497}.

\bibitem{topaz}G.~Montagna, O.~Nicrosini, G.~Passarino 
and F.~Piccinini,
``{\tt TOPAZ0~2.0} - A program for computing deconvoluted and realistic
observables around the $Z^0$ peak'', CERN-TH.7463/94, in press 
on Comput. Phys.
Commun.; G.~Montagna, O.~Nicrosini, G.~Passarino, F.~Piccinini 
and R.~Pittau,
Comput. Phys. Commun. { 76} (1993) 328.  

\bibitem{tressi} This program $TRESSI$ is available, 
upon request, from the authors of this paper.
%
\end{thebibliography}
\end{document}